\documentclass[pra,aps,preprint]{revtex4}

 \usepackage{graphicx}% Include figure files
 \usepackage{dcolumn}% Align table columns on decimal point
 \usepackage{bm}% bold math
 \usepackage{subfigure}
 \newcommand{\YSO}{Y$_2$SiO$_5$}
 \newcommand{\ket}[1]{\left|#1\right\rangle} 
 \newcommand{\bra}[1]{\left\langle#1\right|} 
 \newcommand{\mean}[1]{\left\langle#1\right\rangle}

\begin{document}

 \title{Experimental demonstration of quantum
state tomography  and qubit-qubit interactions for rare-earth-ion based solid
state qubits}

 \author{J. J. Longdell}
 \email{jevon.longdell@anu.edu.au}
 \author{M. J. Sellars}
 \affiliation{Laser Physics Centre, Research School of Physical
    Sciences and Engineering, Australian National University,
    Canberra, ACT 0200, Australia.}%Lines break automatically or can be forced with \\ 
  \date{August 29, 2002}% It is always \today, today,
               %  But any date may be explicitly specified
\begin{abstract}
  We report on the implementation of quantum state tomography for an
  ensemble of Eu$^{3+}$ dopant ions in a \YSO\ crystal. The tomography
  was applied to a qubit based on one of the ion's optical transitions.
  The qubit was manipulated using optical pulses and measurements were
  made by observing the optical free induction in a phase sensitive
  manner. Fidelities of $>90$\% for the combined preparation and
  measurement process were achieved.  Interactions between the ions
  due to the change in the ions' permanent electric dipole moment when
  excited optically were also measured.  In light of these results, the
  ability to do multi-qubit quantum computation using this system is
  discussed.
\end{abstract}

  % PACS, the Physics and Astronomy
                               % Classification Scheme.
  %Use showkeys class option if keyword
                                %display desired
\maketitle

\section{\label{sec:intro}Introduction}
  
Rare earth ions doped into inorganic crystals are promising candidates
for demonstrating quantum computing operations
\cite{pryd00,ichi01,ohls02}. In such a computer it is envisioned that
the quantum information would be stored in the nuclear states of the
ions. These can have life times of many hours \cite{yano91} and
coherence times have been observed as long as 80 ms \cite{elliot}. However,
the single and multi-qubit operations would be carried out optically.
The single qubit operations are possible because of long coherence
times for the optical transitions \cite{ultraslow}. These long
coherence times result in homogeneous linewidths which are much smaller
than both the hyperfine splitting and the available Rabi frequencies
at which the transitions can be driven.  Further to this the large
ratio of inhomogeneous (typically GHz) to homogeneous (sub kHz)
linewidths allow many different qubits to be addressed.  The
multi-qubit operations would be carried out using the strong electric
dipole-dipole interaction between the ions' optical transitions. This
interaction is due to the fact that the ions have a permanent electric
dipole moment which is different for the ground and excited states
\cite{huan89,graf98}.  This interaction has the advantage that quantum
information transferred to the nuclear transitions is
insensitive to such interactions \cite{anabelsthesis}

 The nuclear state of the ions can be initialised to
very high fidelity using optical pumping. 

Here we demonstrate the ability to address a spectrally narrow subset of the
ions. We place their states at any given point on the Bloch sphere and
then read out their states using a sequence of optical pulses,
thus demonstrating their utility as qubits. The interaction
between these qubits due to the electric dipole-dipole
interactions \cite{huan89,graf98} between the ions is measured. From
measurements of these interactions we determine the conditions required
to demonstrate two qubit operations. 

\section{Experimental} 
 
The setup used was similar to that used in \cite{pryd00}. A highly
stabilized dye laser was used with an established stability of better
than 200~Hz over timescales of 0.2~s. The light incident on the sample
was gated with two acousto-optic modulators (AOMs) in series. These
allowed pulses with an arbitrary amplitude and phase envelope to be
applied to the sample. The overall frequency shift introduced by the
AOMs was 10~MHz. The transmitted light was combined with a laser beam
unshifted in frequency in a Mach-Zehnder interferometer. The beat
signal detected with a photo-diode was combined in phase and
quadrature with a 10 MHz reference to allow phase sensitive detection.
In addition to the setup of \cite{pryd00} an auxiliary optical beam
and radio frequency (RF) fields were used to repump the ground state
hyperfine levels. The measurements were carried out with the crystal
in the temperature range \mbox{3-5~K.}

The \YSO\ sample had 0.5~at.\% europium, which occupies the yttrium
sites in the lattice. There are two stable isotopes of europium of
near equal abundance and there are two different crystallographic
sites for yttrium in the lattice. The experiment was performed on
the $^{151}$Eu ions that substituted for yttrium at ``site 1'' \cite{yano91}.

Applying optical pulses of definite area requires that all the ions
experience the same generalized Rabi frequency.  This requirement is
problematic for a number of reasons. It requires the spread in
resonance frequencies of the ions (c. 10~GHz for the whole sample) to
be much smaller than the on-resonance Rabi frequency (c. 500~kHz
available with 200~mW of laser power). It also requires that all ions
have the same oscillator strength, which means that the resonant
transition must be from the same ground state hyperfine level to the
same excited state hyperfine level for all the ions. Further to these
it also requires illuminating all the ions with the same intensity of
light rather than the normal Gaussian laser beam profile.

\begin{figure}
%   \centering
%   \input{level_diag.eepic}
%\ifx\pdftexversion\undefined
  \includegraphics[width=0.2\textwidth]{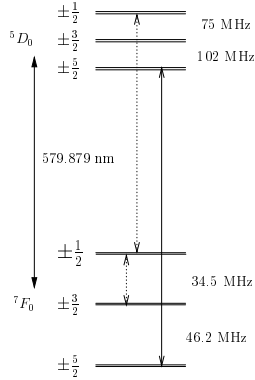}
%\else
%   \includegraphics[width=0.2\textwidth]{level_diag.pdf}
%\fi
   \caption{Energy level diagram for $^{151}$Eu at site 1.
     The experiment was performed on the $\pm 5/2 \rightarrow \pm 5/2$
     transition. The RF and optical fields used for burning back are
     shown with dotted lines.} 
   \label{fig:energylevels}
\end{figure}

Optical holeburning techniques similar to those introduced by
Pryde~et.~al.\cite{pryd00} were used to isolate a subset of ions to
which accurate area pulses could be applied.  All the ions that were
not wanted in the subset of ions and that had a resonance close to the
laser were optically pumped to another hyperfine level and thus far
from resonance with the laser.  The procedure was as follows: First
all ions with a resonance close to the laser frequency were optically
pumped to another hyperfine level.  Then one RF field and the
auxiliary optical beam was applied to the sample. The RF field was
swept in frequency $34.5\pm1.0$~MHz and the auxiliary optical beam was
shifted from the laser frequency by 95.9~MHz. The effect was to repump
the $^{151}$Eu ions with the $\pm 5/2 \rightarrow \pm 5/2$ transition
frequency close to that of the laser into the $\pm 5/2$ ground state.
These transitions are shown on the energy level diagram,
FIG.~\ref{fig:energylevels}, the hyperfine splittings given are those
measured by Yano et al. \cite{yano91} and from such splittings one would
calculate a offset for the auxiliary beam of 96.3~MHz rather than
95.9~MHz. The reason for this discrepancy is the $\pm 1$~MHz
uncertainty in the excited state hyperfine splittings. The value of
95.9~MHz was obtained by adjusting the offset until the absorbative
feature that it created was centered on the laser frequency.

This resulted in a 300~kHz wide feature with a Lorentzian-like
lineshape. This was
then narrowed to a 50~kHz feature with a rectangular lineshape by
applying `zero area pulses' to the ions. The laser beam is modulated
with an envelope given by the difference of two sinc ($\sin(\pi
x)/(\pi x)$) functions.  The lengths and amplitudes of the two sinc
functions were chosen to so that the resulting pulse excited all the ions with a transition
frequency within 500~kHz of the laser except those within 25~kHz
of the laser.

The spread in the optical frequencies of the resulting subset was then narrow
compared to the available resonant Rabi frequencies, however because of the
spatial distribution of intensity across the laser beam we do not
yet have an ensemble to which we can apply precise area pulses.

To select out ions in a particular region of the laser beam, a series
of 4~$\mu$s long pulses were applied to the sample. Some of the ions
experienced a $2\pi$ pulse and were left in the ground state while ions
that saw a different intensity were pumped to a different hyperfine
level.  Ten $2\pi$ pulses were applied with an 8~ms delay between them
to allow the excited ions to decay. Afterwards it could be concluded
from the free induction decays for various pulse lengths that the Rabi
frequency spread of the ensemble was of the order of 10\%.
%\section{Quantum state tomography }

For pedagogical reasons it is helpful to define an explicit phase
reference. The
phase reference for the experiment was given by the light that would
hit the sample if there was no phase shift applied in the RF drives.
The pulse sequence used for the tomography consisted of three pulses
each separated by 70~$\mu$s. First a pulse of an arbitrary length and
with arbitrary phase relative to the phase reference. This was
followed by a 2~$\mu$s long $\pi$ pulse in phase with the phase
reference, to rephase the inhomogeneous broadening in the sample. At
$t=140\mu$s a 1~$\mu$s long $\pi/2$ pulse was applied that was also in
phase with the phase reference. The axes for the Bloch vectors were
chosen such that the ground state ($\ket{0}$) was along the negative
$z$ axis and laser pulses that were in phase with the phase reference caused
rotations about the $y$ axis.

The first pulse was used to create an arbitrary state, and the
coherent emission resulting from this and the rest of the sequence
constituted a measurement of this state. The first pulse was therefore
viewed as causing a rotation of the state vector on the Bloch sphere
and the following two pulses viewed as a rotation of both the
state vector and the Bloch sphere itself.  At each point the coherent
emission measures the projection of the state onto the horizontal
plane.  Thus the amplitude of the component of the emission after the
first pulse that was in phase (in quadrature) with the phase reference was
proportional to $\mean{-X}$ ($\mean{Y}$) for the initial state. This
coherent emission decayed over $\approx 20 \mu \textrm{s}$ (1/(50
kHz)) due to the inhomogeneous linewidth of the ensemble. The ensemble
rephased at 140~$\mu$s and at that time a 1 $\mu$s long $\pi/2$ pulse
was applied.  As the ensemble was rephasing before the $\pi/2$ pulse
the amplitude of the emission in phase (in quadrature) with the phase reference
was proportional to $\mean{X}$ ($\mean{Y}$) for the initial state.
After the $\pi/2$ pulse the amplitude of the emission in phase (in
quadrature) with the phase reference  was proportional to $\mean{-Z}$
($\mean{Y}$) for the initial state. These measurements over-determine
the initial Bloch vector, and linear least squares was used to extract
the measured Bloch vector. The length of the Bloch vector was
calibrated using two particular input states:
$(\ket{0}+\ket{1})/\sqrt{2}$ prepared using a $\pi/2$ pulse with zero
phase shift and the state $\ket{0}$ which required no preparation.%prepared by not having
                                %a preparation pulse. 

\begin{figure}
%\ifx\pdftexversion\undefined
  \subfigure[]{\label{fig:bloch1}\includegraphics[width=0.146\textwidth]
    {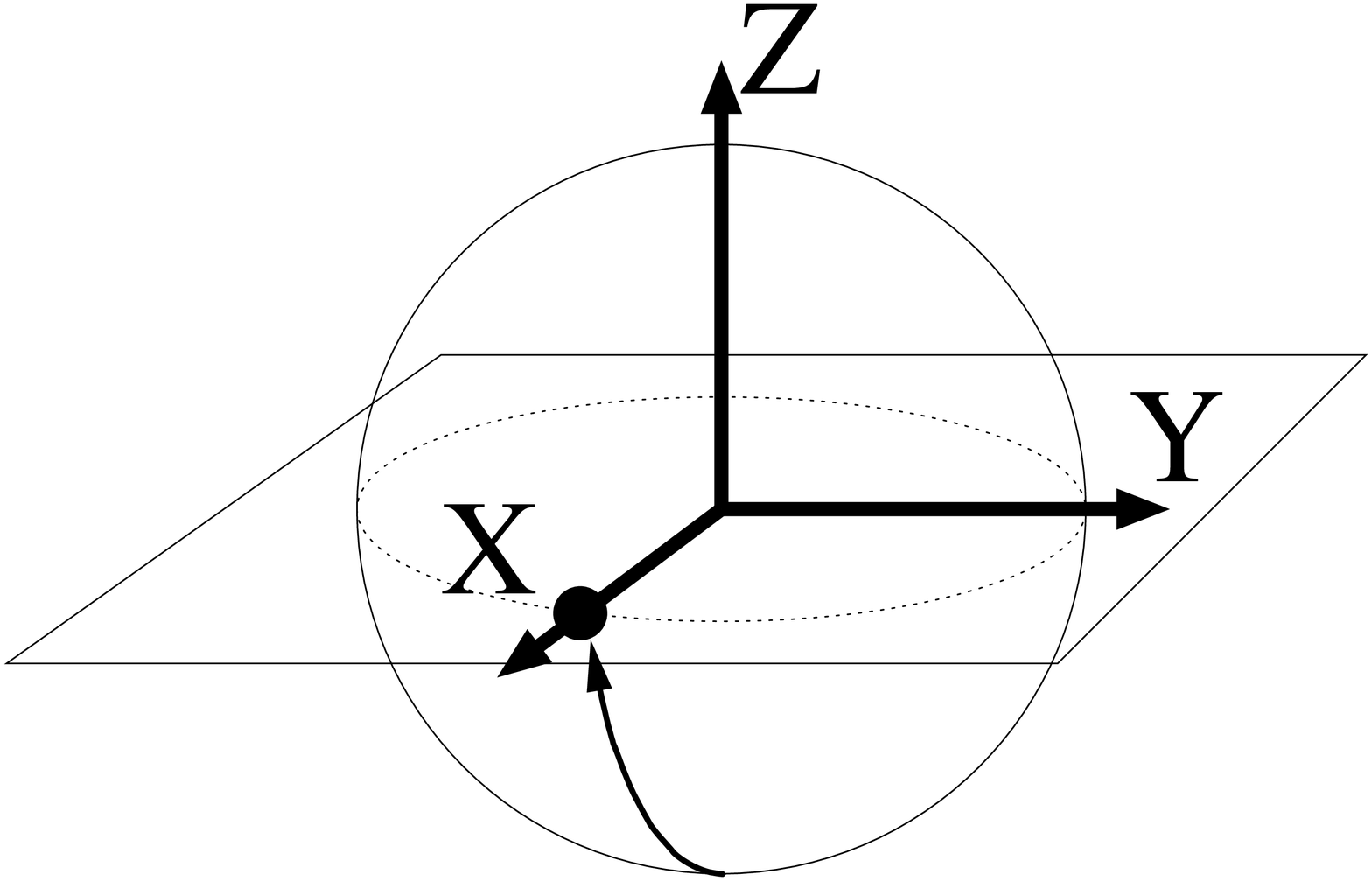}}
  \subfigure[]{\label{fig:bloch2}\includegraphics[width=0.146\textwidth]
    {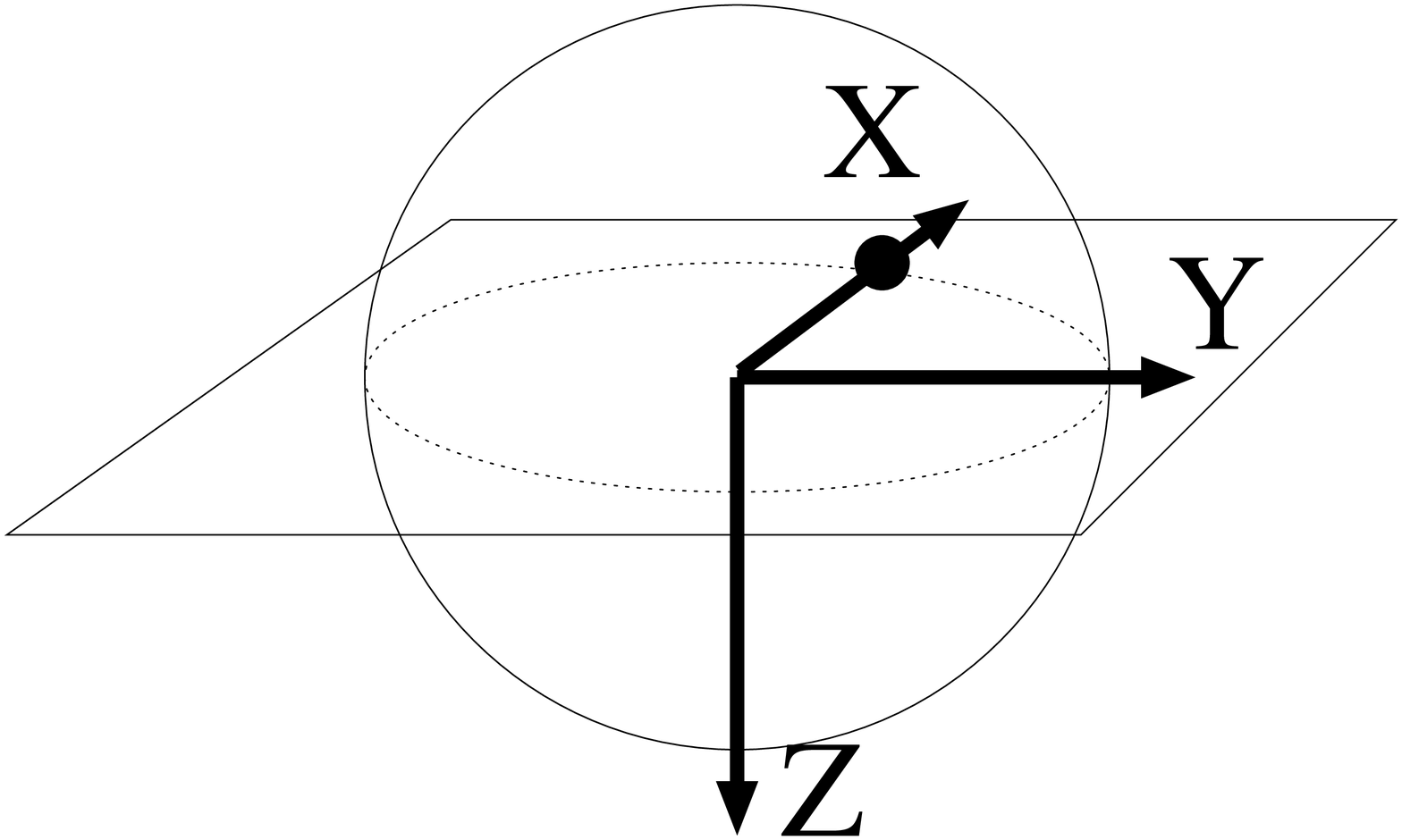}}
  \subfigure[]{\label{fig:bloch3}\includegraphics[width=0.146\textwidth]
    {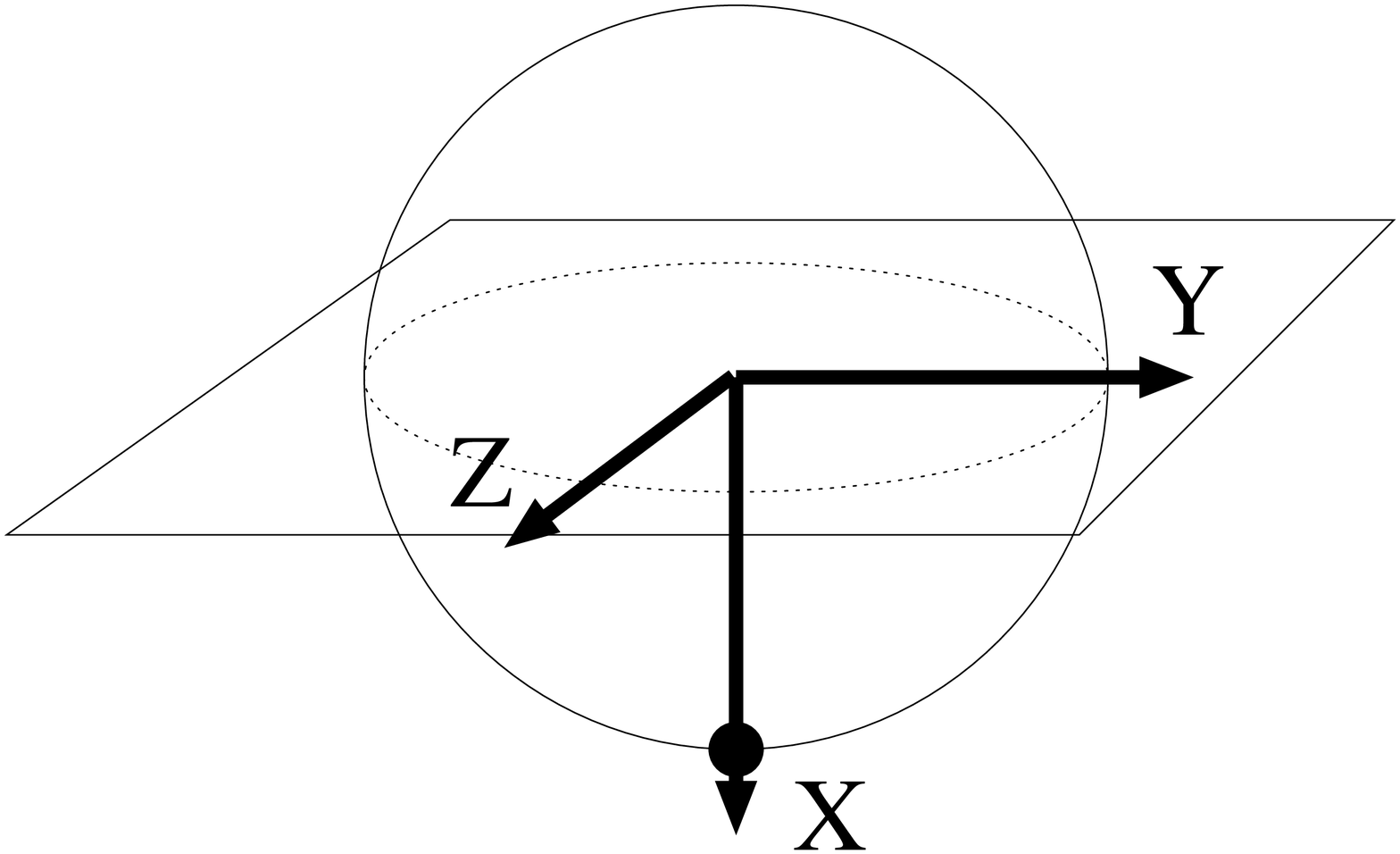}}
%\else
%   \subfigure[]{\label{fig:bloch1}\includegraphics[width=0.146\textwidth]
%    {bloch1.pdf}}
%  \subfigure[]{\label{fig:bloch2}\includegraphics[width=0.146\textwidth]
%    {bloch2.pdf}}
%  \subfigure[]{\label{fig:bloch3}\includegraphics[width=0.146\textwidth]
%    {bloch3.pdf}}
%\fi
  \caption{\label{fig:balls}
    The effect of the tomography sequence on the Bloch sphere for the
    case where the preparation pulse is a $\pi/2$ pulse that is
    in phase with the phase reference. This pulse was to create the
    $(\ket{0}+\ket{1})/sqrt{2}$, the coherent emission from this pulse
    and from the following two pulses constituted the tomography. For
    the case of a general initial state, the axes on the figures would
    be the same but the dot would be in a different place relative to them. 
    Experimental
    results for a the particular shot we are considering here are shown in
    FIG.~\ref{fig:oneshot}.
    (a) The state is moved to the equator by
    the first pulse in the sequence. This gives coherent emission out of phase
    with the laser. (b) Just before $t=140\mu$s the system rephases on
    the opposite side of the Bloch sphere, giving coherent emission in
    phase with the laser.(c) The final $\pi/2$ pulse rotates the state
    to the ground state, which results in no coherent emission.
  }
\end{figure}

 The fidelity is given by 
$  \mathcal{F} = \bra{\phi}\rho\ket{\phi}$,
where $\ket{\phi}$ is the input state and $\rho$ is the measured
density matrix.
A significant contribution to the error in the tomography process was
the shot to shot variation in the number of ions to which the process
was applied.  The requirement of phase coherence for the 200 $\mu$s of
the tomography sequence was easily satisfied by the laser.  However
the preparation of the spike took approximately 10 s, leading to the
possibility that the laser drifts a significant fraction of the 50 kHz
width of the spike in this time. This caused a shot to shot variation
in the number of ions to which the tomography was applied which in turn
had the effect of scaling the length of the measured Bloch vector. If
a state that is being measured can be assumed to be pure, the measured
Bloch should be normalized. This normalization makes the tomography
process insensitive to drifts in the number of ions to which the
process is applied. This leaves the inhomogeneity in the Rabi frequencies as the main source of error in the process.

The fidelity of the combined state preparation and tomography are
shown in table \ref{tab:results}. 

\begin{figure}
%\ifx\pdftexversion\undefined
  \includegraphics[width=0.37\textwidth]{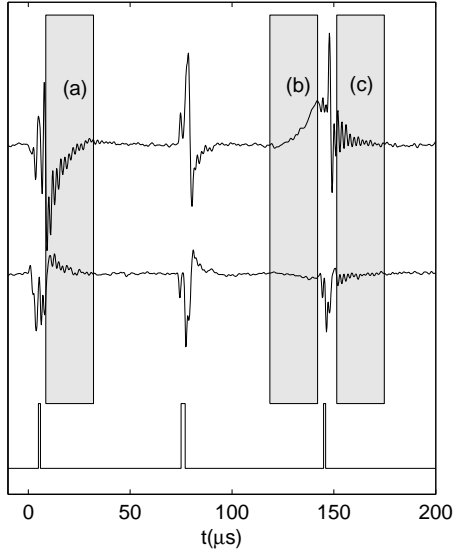}
%\else
%   \includegraphics[width=0.39\textwidth]{oneshot2f.pdf}
%\fi
  \caption{The results of a quantum state
    tomography sequence. In the top trace the amplitude of the
    coherent emission in phase with the applied light is plotted
    versus time. The second trace shows the component in quadrature
    with the applied light. The applied optical pulses saturate the
    detector which takes $\approx 10 \mu$s to recover. The positions
    of the applied light pulses as shown schematically in the third
    trace. The labels (a), (b) and (c) correspond to the snapshots of the
    Bloch spheres shown in FIG.~\ref{fig:balls}. The regions over
    which the signal was integrated to arrive at measurements are
    shown in grey.
  \label{fig:oneshot}}
\end{figure}

\begin{table}
  \begin{ruledtabular}
  \caption{Fidelity of the combined state preparation and tomography
    for different input states. Each point was repeated three times
    and the fidelities reported are the worst of those repeats.  
     The position of the last state on the Bloch sphere corresponds to
     the position of Canberra on the Earth.
% See
%    the text for a description of fidelity2.% The  other two
%    values for  fidelity2 for 
%    the north pole were both 0.99
 \label{tab:results}}
  \begin{tabular}{cccc}
    State   & Fidelity & Fidelity \\
       &             & assuming \\
       &             &pure state \\
    $(\ket{0}+i\ket{1})/\sqrt{2}$ &  0.95 & 0.95\\
    $(\ket{0}+\ket{1})/\sqrt{2}$ &  0.95 & 0.95\\
    $\ket{0}$ &  0.96 & 0.97\\
    $(\ket{0}-\ket{1})/\sqrt{2}$ &  0.96 & 0.97\\
    $\ket{1}$         &  0.88 & 0.89\\
    $(\ket{0}-i\ket{1})/\sqrt{2}$  &  0.88 & 0.96\\
   $\cos(0.960)\ket{0} + \sin(0.960)\exp(2.60i)\ket{1}$ &  0.81 & 0.99\\
  \end{tabular}
  \end{ruledtabular}
\end{table}

The results of one particular experimental shot are shown in
FIG.~\ref{fig:oneshot}.  Here the state to be measured is created at
$t=0\mu$s with a $\pi/2$ pulse unshifted in phase. This puts the Bloch
vector along the $x$ axis. This gave coherent emission out of phase with
the laser, which decayed as the inhomogeneous broadening dephased the 
ensemble. At $t=70\mu$s the $\pi$ pulse was applied to rephase the
inhomogeneous broadening. This should have caused no polarisation of
the ions and thus produced no coherent emission. As $t=140\mu$s
approaches the ensemble rephased on the opposite side of the Bloch
sphere producing coherent emission in phase with the laser.  At
$t=140\mu$s a $\pi/2$ pulse was applied which takes the ions down to
the ground state, stopping the coherent emission. The high frequency
ringing superimposed on the emission following the application of the
pulses is due to the pulses exciting the edges of the trenches in which the
spikes lie.

\section{The coherent interaction between two ensembles}
\begin{figure}
%\ifx\pdftexversion\undefined
    \includegraphics[width=0.4\textwidth]{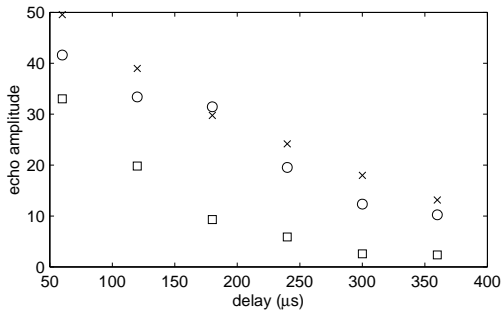}
%\else
%     \includegraphics[width=0.4\textwidth]{demol.pdf}
%\fi
  \caption{A photon echo sequence was applied to a spike. During the
    sequence a perturbing pulse is applied to another spike prepared 5
    MHz away. The echo amplitude vs delay between the two applied
    pulses. The crosses correspond to no perturbing pulse, the circles
    denote a perturbing pulse applied just after the $\pi/2$ pulse and the
    squares just after the $\pi$ pulse.}
  \label{fig:demol}
\end{figure}

Interactions between the europium ions were also measured. Each europium
ion in Eu:\YSO\ has a permanent electric dipole moment which changes
by a few percent when the ion is excited optically. A europium ion
will see a change in electric field and hence its resonant frequency
changes when a nearby ion is excited \cite{huan89,graf98}.

This interaction is large for ions that are close, however when working
with two ensembles each of which constitute a small fraction of the
inhomogeneous line, the distance from a particular member of one
ensemble to any member of the other is much larger than the crystal's
mean inter-dopant separation. Because of the random nature of the
relative positions of a given pair of ions, the excitation of one
ensemble causes a random shift in the resonance frequencies of each of
the members of the other.

In order to measure the frequency shifts in one ensemble due to the
excitation of another the approach of Huang~et~al. is used
\cite{huan89}. A photon echo sequence is applied to a spike.
During the sequence a perturbing pulse is applied to another spike
prepared 5~MHz away. When the perturbing pulse was applied just after
the $\pi$ pulse the amplitude of the echo was reduced significantly.
When the perturbing pulse was applied just after the $\pi/2$ pulse the
reduction in the amplitude of the echo was much smaller. 
This confirms that the perturbing pulse induces a frequency shift
which, if applied at the start of the sequence, is rephased. This
result shows that the frequency shift caused by exciting the second
qubit is constant over a time which is  approximately that of the
excited state lifetime.

From FIG.~\ref{fig:demol} we can see that the excitation of the perturbing
ensemble causes a spread of $\approx 1$ kHz ($1/(2\pi\times200\mu$ s$)$)
in the resonant frequencies.  The ensemble to which the perturbing
pulses were applied was prepared by burning a trench and then burning
back an antihole. This $\approx 300$ kHz wide feature has
approximately $10^{-6}$ of the ions.
This corresponds to a mean inter-dopant separation 100 times greater
than for the full ensemble or $100\times 2.5=250$ nm.  As we can
expect the interaction to scale as $r^{-3}$, ions separated by the
mean inter-europium distance in this sample can be expected to see
shifts of $\approx 1$ GHz.

%\section{Discussion}

In order to do multi-qubit quantum computation, in addition to the
other forms of inhomogeneity that were overcome in this work, one
would have to overcome the problem of inhomogeneity in ion-ion
interaction strength.  When working with two spikes in the
inhomogeneous line the mean shift of frequencies in one ensemble is a
small percentage of the other's inhomogeneous linewidth. This makes the
method of Ohlsson et.~al. \cite{ohls02} for overcoming interaction
strength inhomogeneity very difficult to implement as it relies on
selecting ions pairs with frequency shifts greater than the
inhomogeneous linewidth of the ensembles. 

%  However if one has the
% ability to perform accurate single qubit operations on the ensembles
% the interactions need only be large compared to the the homogeneous
% linewidths. The pulse sequence would be similar to that used to
% measure the interactions except at the end of the sequence it all the
% ions that have experienced a particular phase shift due to the
% interaction would be taken back to the ground state, leaving the
% others to some extent excited. Repeated application of such a pulse
% sequence would optically pump all the ions from the ensemble except
% those with the desired interaction strength. While this would leave
% ensembles with much more ions than would the Ohlsson scheme the number
% of ions left however would still decrease rapidly with the number of
% qubits making such an approach not scalable.

% The problem of inhomogeneity of ion-ion interaction strength could be
% overcome using ``molecules'', that is a large number of identical
% ensembles, in the same way as in liquid state NMR quantum computing.
% At present however, no such materials present themselves.  The
% detection of single ions would also solve this problem. The optical
% detection of single impurity sites in solids has been demonstrated
% \cite{grub97}, although the low oscillator strengths for rare earths
% and the requirement that the detection be spin selective would make
% the task more difficult. In the authors' opinion this compares
% favorably with the situation for other schemes for the detection of
% single qubits in solids.

\section{Conclusion}

In conclusion we have demonstrated quantum state tomography on
Eu$^{3+}$ dopant ions in \YSO, with fidelities of typically $>90$\%
achieved.  We also measured the frequency shifts of one ensemble of
dopant ions caused by the excitation of another group of dopant ions.
%
%
% The detection of the spin state of single ions would solve the problem of
% inhomogeneity in the ensembles enabling  multi quantum computing, with
% significantly higher fidelities than demonstrated here.
%
% While scalable quantum computing using ensembles selected from the
% inhomogeneous line will not be possible, demonstrations for small
% numbers of qubits are feasible. 
The results of these measurements suggest that if we can achieve high
fidelity single qubit operations on ensembles $\approx$300~kHz wide 
then multi-qubit operations will be possible.

We were restricted to working with Rabi frequencies of 250~kHz by the
bandwidth of the modulation system used to apply the pulses and by the
available laser power. Intensities of 200~W/cm$^2$ are needed for
250~kHz Rabi frequencies. It is envisioned that the use of moderately
higher laser powers, smaller spot sizes, improvements to the
modulation system and perhaps the use of composite pulse sequences
will allow the use of ensembles with this larger spectral width. The
larger average interaction strengths associated with this should allow
the demonstration of two qubit operations for ensembles of ions.

The authors would like to thank Geoff Pryde and Neil Manson for their
helpful comments on the manuscript.

%\bibliography{../refs/refs.bib}

\begin{thebibliography}{9}
\expandafter\ifx\csname natexlab\endcsname\relax\def\natexlab#1{#1}\fi
\expandafter\ifx\csname bibnamefont\endcsname\relax
  \def\bibnamefont#1{#1}\fi
\expandafter\ifx\csname bibfnamefont\endcsname\relax
  \def\bibfnamefont#1{#1}\fi
\expandafter\ifx\csname citenamefont\endcsname\relax
  \def\citenamefont#1{#1}\fi
\expandafter\ifx\csname url\endcsname\relax
  \def\url#1{\texttt{#1}}\fi
\expandafter\ifx\csname urlprefix\endcsname\relax\def\urlprefix{URL }\fi
\providecommand{\bibinfo}[2]{#2}
\providecommand{\eprint}[2][]{\url{#2}}

\bibitem[{\citenamefont{Pryde et~al.}(2000)\citenamefont{Pryde, Sellars, and
  Manson}}]{pryd00}
\bibinfo{author}{\bibfnamefont{G.~J.} \bibnamefont{Pryde}},
  \bibinfo{author}{\bibfnamefont{M.~J.} \bibnamefont{Sellars}},
  \bibnamefont{and} \bibinfo{author}{\bibfnamefont{N.}~\bibnamefont{Manson}},
  \bibinfo{journal}{Phys. Rev. Lett.} \textbf{\bibinfo{volume}{84}},
  \bibinfo{pages}{1152} (\bibinfo{year}{2000}).

\bibitem[{\citenamefont{Ichimura}(2001)}]{ichi01}
\bibinfo{author}{\bibfnamefont{K.}~\bibnamefont{Ichimura}},
  \bibinfo{journal}{Optics Comm.} \textbf{\bibinfo{volume}{196}},
  \bibinfo{pages}{119} (\bibinfo{year}{2001}).

\bibitem[{\citenamefont{Ohlsson et~al.}(2002)\citenamefont{Ohlsson, Mohan, and
  Kr\"oll}}]{ohls02}
\bibinfo{author}{\bibfnamefont{N.}~\bibnamefont{Ohlsson}},
  \bibinfo{author}{\bibfnamefont{R.~K.} \bibnamefont{Mohan}}, \bibnamefont{and}
  \bibinfo{author}{\bibfnamefont{S.}~\bibnamefont{Kr\"oll}},
  \bibinfo{journal}{Optics Comm.} \textbf{\bibinfo{volume}{201}},
  \bibinfo{pages}{71} (\bibinfo{year}{2002}).

\bibitem[{\citenamefont{Yano et~al.}(1991)\citenamefont{Yano, Mitsuanga, and
  Uesugi}}]{yano91}
\bibinfo{author}{\bibfnamefont{R.}~\bibnamefont{Yano}},
  \bibinfo{author}{\bibfnamefont{M.}~\bibnamefont{Mitsuanga}},
  \bibnamefont{and} \bibinfo{author}{\bibfnamefont{N.}~\bibnamefont{Uesugi}},
  \bibinfo{journal}{Opt. Lett} \textbf{\bibinfo{volume}{16}},
  \bibinfo{pages}{1884} (\bibinfo{year}{1991}).

\bibitem[{ell()}]{elliot}
\bibinfo{howpublished}{Elliot Fraval, Personal Communication}.

\bibitem[{\citenamefont{Equall et~al.}(1997)\citenamefont{Equall, Sun, Cone,
  and Macfarlane}}]{ultraslow}
\bibinfo{author}{\bibfnamefont{R.~W.} \bibnamefont{Equall}},
  \bibinfo{author}{\bibfnamefont{Y.}~\bibnamefont{Sun}},
  \bibinfo{author}{\bibfnamefont{R.~L.} \bibnamefont{Cone}}, \bibnamefont{and}
  \bibinfo{author}{\bibfnamefont{R.~M.} \bibnamefont{Macfarlane}},
  \bibinfo{journal}{Phys. Rev. Lett.} \textbf{\bibinfo{volume}{72}},
  \bibinfo{pages}{2179} (\bibinfo{year}{1997}).

\bibitem[{\citenamefont{Graf et~al.}(1998)\citenamefont{Graf, Renn, Zumofen,
  and Wild}}]{graf98}
\bibinfo{author}{\bibfnamefont{F.~R.} \bibnamefont{Graf}},
  \bibinfo{author}{\bibfnamefont{A.}~\bibnamefont{Renn}},
  \bibinfo{author}{\bibfnamefont{G.}~\bibnamefont{Zumofen}}, \bibnamefont{and}
  \bibinfo{author}{\bibfnamefont{U.~P.} \bibnamefont{Wild}},
  \bibinfo{journal}{Phys. Rev. B} \textbf{\bibinfo{volume}{58}},
  \bibinfo{pages}{5462} (\bibinfo{year}{1998}).
\bibitem[{\citenamefont{Huang et~al.}(1989)\citenamefont{Huang, Zhang, Lezama,
  and Mossberg}}]{huan89}
\bibinfo{author}{\bibfnamefont{J.}~\bibnamefont{Huang}},
  \bibinfo{author}{\bibfnamefont{J.~M.} \bibnamefont{Zhang}},
  \bibinfo{author}{\bibfnamefont{A.}~\bibnamefont{Lezama}}, \bibnamefont{and}
  \bibinfo{author}{\bibfnamefont{T.~W.} \bibnamefont{Mossberg}},
  \bibinfo{journal}{Phys. Rev. Lett.} \textbf{\bibinfo{volume}{63}},
  \bibinfo{pages}{78} (\bibinfo{year}{1989}).
\bibitem[{\citenamefont{Alexander}()}]{anabelsthesis}
\bibinfo{author}{\bibfnamefont{A.~L.} \bibnamefont{Alexander}},
  \bibinfo{note}{honours Thesis, Australian National University,
  http://eprints.anu.edu.au/archive/00000761/}.
\end{thebibliography}

\end{document}